# A Family Group of Two-dimensional Node-Line Semimetals

Yuanjun Jin,[†] Rui Wang,[†,‡] Jinzhu Zhao,[†,§] Candi Zheng,[†] Li-Yong. Gan,[†,//] Junfeng Liu,[†] Hu Xu,[†,*] and S. Y. Tong.[†,⊥,*]

[†]Department of Physics, South University of Science and Technology of China, 518055 Shenzhen, China

[‡]Institute for Structure and Function & Department of Physics, Chongqing University, 400030 Chongqing, China

[§]Dalian Institute of Chemical Physics, Chinese Academy of Sciences, 116023 Dalian, China

[//]Key Laboratory of Advanced Technology of Materials (Ministry of Education), Superconductivity and New Energy R&D Center, Southwest Jiaotong University, Chengdu, 610031 Sichuan, China

[⊥]School of Science and Engineering, The Chinese University of Hong Kong (Shenzhen), 518172 Shenzhen, China

**ABSTRACT:** Using evolutionary algorithm and first-principles calculations, we predict a family group of two-dimensional node-line semimetals MX (M=Pd, Pt; X=S, Se, Te), which has zig-zag type mono-layer structure in *Pmm2* layer group. Band structure analysis reveals that node-line features are caused by band inversion and the inversion exists even in the absence of spin-orbital-coupling. Tests are carried out to confirm that the node-line loop is protected by crystal symmetry. This work extends our knowledge of node-line materials to two-dimensional cases, i.e., a group of metal-group VI compounds sharing the same lattice structure which has time reversion and crystal-mirror inversion symmetries.

Since the discovery of topological insulators (TIs) and topological semimetals (TSMs), our knowledge to distinguish between materials extends beyond conventional standards such as investigating the existence or amplitude of electronic band gaps, but has to include the topology of band structures in momentum space.[1-4] Physical properties such as the dispersion of surface states connecting valence and conduction bands depend on the global topology, and not on local details. More recently, the concept has been expanded to include a broader notion: symmetry-protected topological (SPT) phases,[5,6] of which TI is just an example. In the broader notion, the topology of a SPT phase depends on the presence of a symmetry, e.g., TIs are protected by the time-reversal (TR) symmetry while topological crystalline insulators (TCIs)[7-14] are protected by the crystal symmetry (CS).

So far, three types of topological semimetals have been proposed. These are: the Dirac type,[15-18] Wely type[19-22] and node-line (NL) type.[23-25] Only 3D materials have been reported to-date to host node-line type band structures,[26-28] e.g. the anti-perovskite compounds.[29,30] In this letter, we propose that a family of compounds with zig-zag mono-layer MX structure, in the layer group *Pmm2* (No.23), hosts 2D node-line semimetal features. In the compounds, M refers to a transition metal atom having d orbitals, e.g., Pd and Pt, while X refers to a group VI element atom having p orbitals, e.g., S, Se and Te. Due to strong common features shared by the compounds, the paper will focus mainly on the properties of PdS. Detailed data beyond PdS will be supplied in the Supplementary Information (SI).

The structure of the mono-layer MX compound is found by search using the *ab initio* evolutionary algorithm USPEX code.[31,32] The search criterion is minimization of the total energy of the system. Structural optimization, total energy and electronic property calculations are carried out in the framework of density functional theory (DFT) as implemented in the Vienna Ab initio Simulation Package (VASP).[33,34] Potentials based on the projector augmented wave (PAW) method[35] are used to describe ion-electron interactions. The Perdew-Burke-Ernzerhof (PBE)[36] functional is employed. A thin slab is chosen separated by 15 Å vacuum regions along the z direction. Such a slab is found to be numerically well converged. The electronic wave functions are expanded in a plane-wave basis truncated at an energy cutoff of 500 eV. Integrals over the Brillouin zone (BZ) are replaced by sums over k-points equivalent to a two-dimensional (9×15) sampling for the simple p(1×1) phase. Vibrational properties are calculated using the frozen phonon method and generated with the PHONOPY[37] package.

The structure of a mono-layer MX compound is shown schematically in Figure 1. Each mono-layer consists of three atomic planes, with two M atoms at the middle plane and two X atoms at 'top' and 'bottom' planes respectively, forming a 2D zig-zag structure (see Figure 1(c), side view). The lattice constant, M-X bond length, X-M-X bond angle and mono-layer thickness are listed in Table 1 for each compound. Every M atom bonds to four X atoms and vice versa. There are two mirror-inversion planes in the zig-zag MX structure: the *x-z* plane through an X atom and *y-z* plane through either an M or X atom, marked by $m_x$ and $m_y$ respectively in Figure 1. The mirror-inversion symmetry is equivalent to a $C_2$ rotation, with rotation axis along the *z* direction through an X atom, plus central inversion, with inversion point at an X atom.

DFT calculations, listed in Table 1, show that the mono-layer MX structures have very low surface energies, e.g, the 2D PdS structure has a surface energy of 0.32 *J/m²*, significantly lower than that of 1.13 *J/m²* in bulk PdS, which has the *P42/m* space group symmetry. In addition, the calculated vibrational frequencies show no imaginary values in the BZ for these compounds (see Figure S1), indicating their robust lattice stability. In the absence of spin-orbit-coupling (SOC), the electronic band structure of mono-layer PdS projected along high symmetry directions shows no energy gap at the Fermi level (see Figure 2(a)). Three crossing points are observed slightly above or below the Fermi level, located along the lines Γ-X, Γ-Y and Γ-M respectively. The conduction and valence bands at each crossing point exhibit almost linear dispersion.

Orbital projected band structure reveals the band inversion behavior between conduction and valence bands at Γ. There, the conduction band is non-degenerate (twofold degenerate if spin is considered) and it is the same for the valence band. The conduction band is dominated by $Pd_{-5d}$ orbitals and the valence band is dominated by $S_{-3p}$ orbitals. At Γ, bonding and anti-bonding states split in energy and the much bigger split for the $S_{-3p}$ orbitals than that of the $Pd_{-5d}$ orbitals causes the band inversion. Specifically, the gap between the split $S_{-3p}$ orbitals is about 4 eV while that of the split $Pd_{-5d}$ orbitals is only 1.6 eV. Thus, the bonding state of $S_{-3p}$ orbitals moves below the anti-bonding state of $Pd_{-5d}$ orbitals, with a 1.5 eV inversion gap. In these results, SOC is neglected. Band inversion thus causes conduction and valence bands to cross continuously and the crossings form a continuous loop in momentum space (see Figure 2(b)). The loop does not lie flat on the Fermi surface but has finite dispersion in the region near the Fermi level. All the mono-layer MX compounds studied in this work show similar band inversion, node-line feature and gap. The results are listed in Table 1 and SI. When SOC is included, a small gap opens up at each intersection of the bands. The mono-layer MX compounds are weak SOC systems. For example, in mono-layer PdS, the values of the gap along Γ-X, Γ-Y and Γ-M are only 5 *meV*, 27 *meV* and 6 *meV*, respectively (see Figure 3(a)). Another effect of SOC is that the twofold degenerate states along the M-X direction split into two separate states while other parts of the band structure remain essentially the same. More figures showing the gaps and state splitting are found in the Figure S2. Figure 2(c) shows the contribution from each atomic orbital to the node-line loop. It shows that where the contribution of each single orbital was projected along the loop. It is shown that Pd $e_g$ orbital dominate the band in the loop region, more than 80%, and the $t_{2g}$ orbital makes limited contribution. Comparing with the in-plane components of S$_{-p}$ orbital (the combined $p_x$ and $p_y$), the contribution of S$_{-pz}$ can be ignored.

In 3D node line materials,[29,30] metallic surface states are created by the presence of the surface. Thus we expect that in a 2D node-line material, metallic edge states may be created by the presence of an edge. To investigate this possibility, we calculated the electronic bands of a nano-ribbon formed by cutting a mono-layer PdS along the x direction in two places. As shown in Figure 4(b), the nanoribbon approximately 130 Å wide has two edges terminated by Pd and S atoms, respectively. The electronic bands of the nano-ribbon, calculated without SOC effects, contain two edge states (see Figure 4(a)). One of the edge states has energy increasing from the middle of the BZ towards the zone boundary. This edge state is derived from S edge atoms. The other edge state has the opposite behavior: its energy decreases from the middle of the BZ towards the zone boundary. This other edge state is derived from Pd edge atoms. Both edge states exhibit quadratic dispersion behavior and they cross the Fermi level at different locations of the BZ, confirming their metallic nature.

As in 3D node-line structures such as the anti-perovskite compounds,[29,30] band inversion in the 2D mono-layer MX compounds is not caused by SOC. Also, the crossing points in the node-line feature are not high symmetry points in the BZ. Only the center of the node-line loop is a high symmetry inversion point. To investigate symmetry related properties of the various features, we perform two tests. In the first test, compressive or tensile (hydrostatic) strain is applied to the mono-layer PdS compound. Here, we simply compress or extend the lattice constant while keeping other atomic coordinates unchanged. This is equivalent to changing the Pd-S bond length while maintaining the other geometric parameters such as the Pd-S-Pd bond angle unchanged. The applied strain is expected to modify the interaction between the S$_{-p}$ and Pd$_{-d}$ orbitals without changing lattice symmetries of the system. As shown in Figure 3(b), the symmetry-preserving hydrostatic strain sensitively tunes the amplitude of the band inversion gap. However, the crossing points of the bands remain almost unchanged. The size of the node-line loop also is not sensitive to the applied strain, even under extremely exaggerated (e.g., +40%) tensile strains. This means that the node-line features are symmetry protected and insensitive to detailed strength of the coupling between bonding *p* and anti-bonding *d* orbitals.

In the second test, we move one of the Pd atoms in the unit cell by a displacement of with equal *x*, *y* and *z* components of 0.1 Å. All the crystal symmetries, e.g., mirror symmetry along *x* and *y* directions, $C_2$ rotation symmetry, etc., disappear consequently. On the other hand, the interaction between orbitals is not expected to be perturbed much because the Pd-S bond lengths and bond angles remain almost the same. New band structures are calculated using the symmetry-broken structure and we see sizable gaps appear at intersections of bands (see Figure 3(c)). Nano-ribbon calculations using the symmetry-broken structure show that the two edge states remain essentially unaffected by the lattice distortion (see Figure 4(c)). Results of the second test confirm the notion that the node-line features are extremely symmetry sensitive; it is indeed rather remarkable that a small distortion that destroys the symmetry but hardly perturbs the bond strength should totally annihilate the node-line features.

In summary, using evolutionary algorithm and ab initio DFT calculations, we discover a family of 2D monolayer MX compounds with *Pmm2* layer group, composed of transition metal elements with valence *d* orbitals and group VI elements with valence *p* orbitals. First-principles calculations show all the studied mono-layer MX compounds host features of 2D node-line topological band structures in momentum space. The node-line loop near the Fermi level is formed by band inversion at the center of the BZ caused by the interaction between transition metal d orbitals and group VI element p orbitals. The band inversion is not due to SOC effects as in Dirac or Wely type semimetals.[15-22] Calculations of nano-ribbons of 2D mono-layer MX compounds reveal one edge state per lattice edge and the edge states exist independent of the crystal symmetry.[38] On the other hand, the node-line features are protected by lattice symmetry, its survival depends sensitively on the existence of the latter. Small distortions that destroy lattice symmetry will annihilate the node-line loop by opening gaps at band crossing points. Surface energy and phonon calculations of the mono-layer MX compounds demonstrate good potential for realizing node-line type 2D materials in the laboratory.

**Table 1. The parameters of zig-zag mono-layer MX compounds including 2D lattice constant (a and b), bond length ($d_{bond}$), bond angles (α and β), layer thickness ($d_{-thick}$). The value of surface energy ($E_{surf.}$), and the band inversion gap ($gap_{BI}$) are also shown.**

|      | Lattice constant (Å) | | $d_{bond}$ (Å) | α (degree) | β (degree) | $d_{-thick}$ (Å) | $E_{surf.}$ ($J/m^2$) | $gap_{BI}$ (eV) |
|------|------|------|------|-------|-------|------|------|-----|
|      | a    | b    |      |       |       |      |      |     |
| PdS  | 5.54 | 3.20 | 2.40 | 96.11 | 70.12 | 2.26 | 0.32 | 1.5 |
| PdSe | 5.62 | 3.20 | 2.52 | 98.08 | 66.79 | 2.58 | 0.19 | 1.0 |
| PtS  | 5.45 | 3.26 | 2.40 | 94.86 | 68.92 | 2.19 | 0.55 | 2.5 |
| PtSe | 5.49 | 3.48 | 2.53 | 95.00 | 65.62 | 2.46 | 0.27 | 2.5 |
| PtTe | 5.52 | 3.75 | 2.69 | 93.25 | 61.81 | 2.70 | 0.30 | 2.4 |

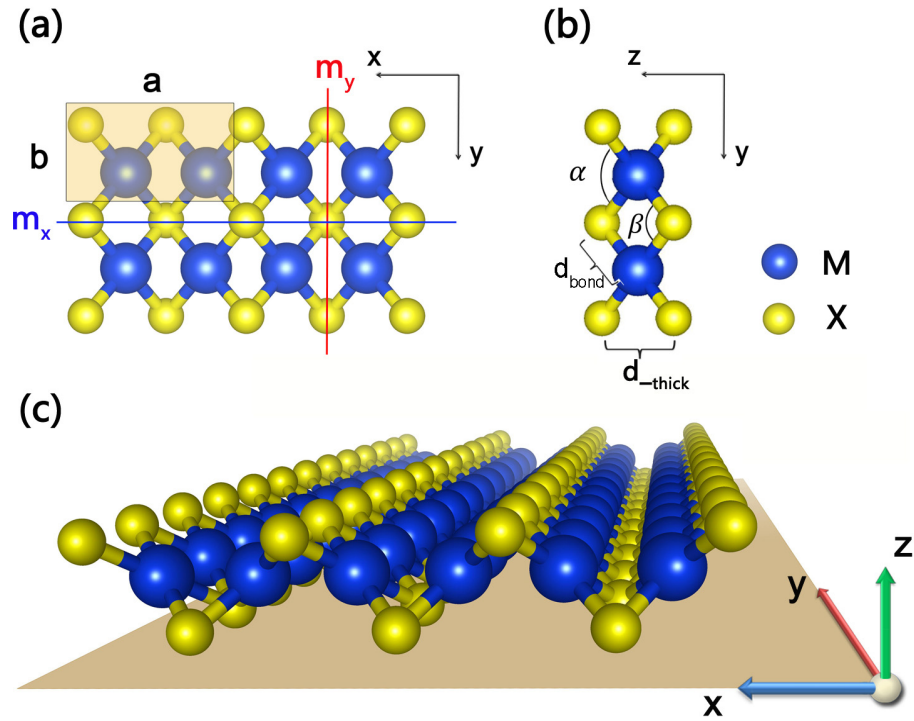

**Figure 1.** The atomic structure of zig-zag mono-layer PdS, where $z$ refers the direction perpendicular to the layer, and $x$ and $y$ are directions parallel to the layer. Top view where the unit cell is highlighted in orange color. In each unit cell, there are two M atoms (blue spheres) and two X atoms (yellow spheres). (b) Side view of the mono-layer PdS viewed in the $x$ direction. Each layer contains three atomic planes parallel to the $x$-$y$ plane. The thickness of the layer $d_{\text{-thick}}$, the M-X bond length $d_{\text{bond}}$, the X-M-X ($\alpha$) and M-X-M ($\beta$) bond angles are marked. (c) A 3D view of the mono-layer PdS crystal.

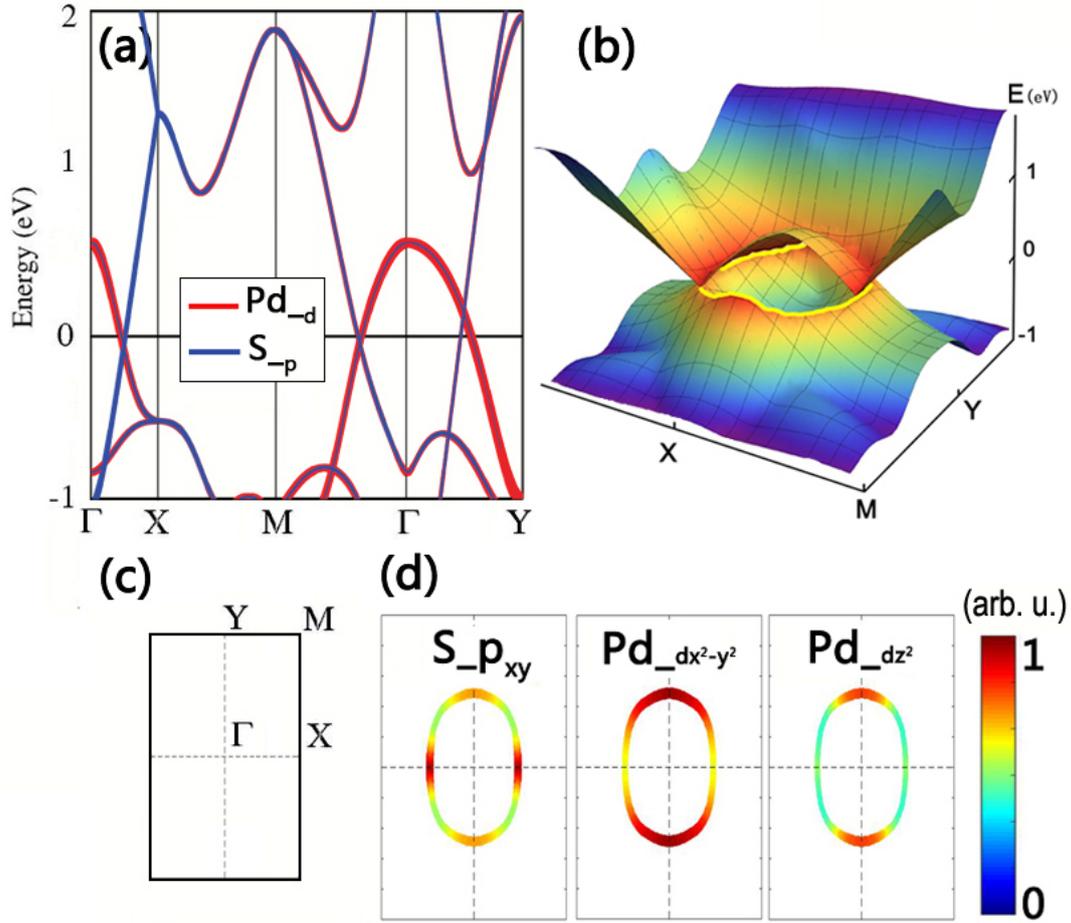

**Figure 2.** The electronic band structure of the zig-zag mono-layer PdS where the Fermi energy is set to zero. (a) Two-dimensional projected fat band structure in the absence of SOC, where the contribution of orbitals are differentiated by colors. Red refers to valence $d$ orbitals of Pd while blue refers to valence $p$ orbitals of S. (b) Three-dimensional band structure of two bands below and above the Fermi level forming a node-line loop due to band crossings. Part of the higher energy band is not shown in order to show the inside region surrounded by the loop (highlighted in yellow). (c) The Brillouin Zone. (d) Contribution, in absolute value, represented by the depth of the color bar (far right), from each valence orbital is shown projected along the path of the loop.

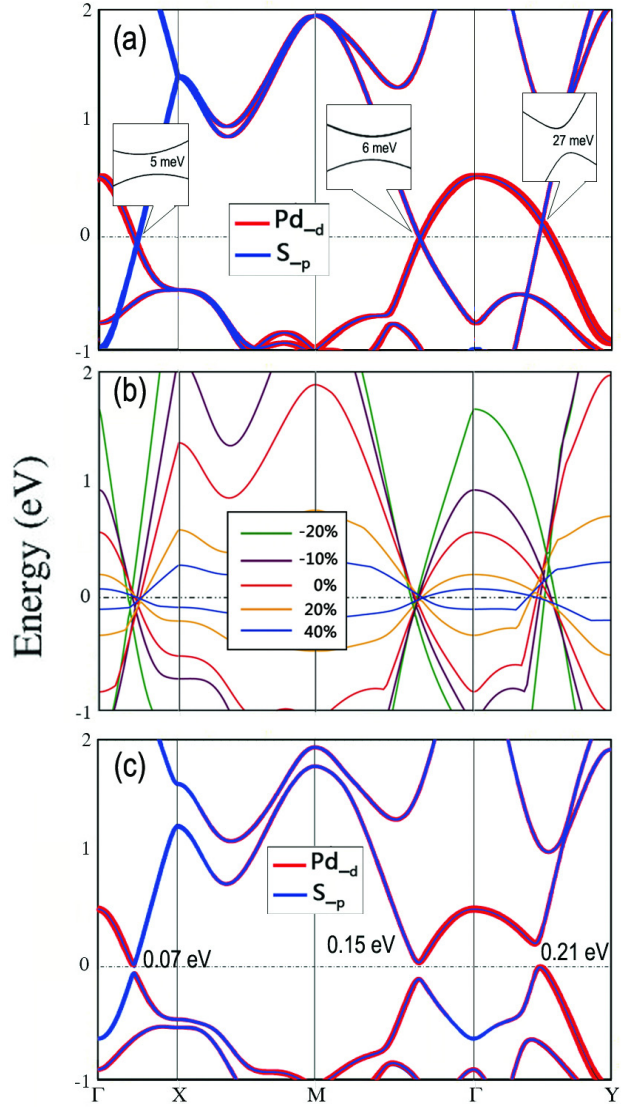

**Figure 3.** Two-dimensional electronic band structure of mono-layer PdS under different conditions. (a) Band structure with SOC. (b) Band structure under different hydrostatic strains. (c) Band structure without SOC where one Pd atom in each unit cell is shifted by a displacement vector whose x, y and z components are 0.1 Å. Red refers to the valence d orbitals of Pd while blue refers to the valence p orbitals of S. Colors in between represent a mixture of red and blue orbitals.

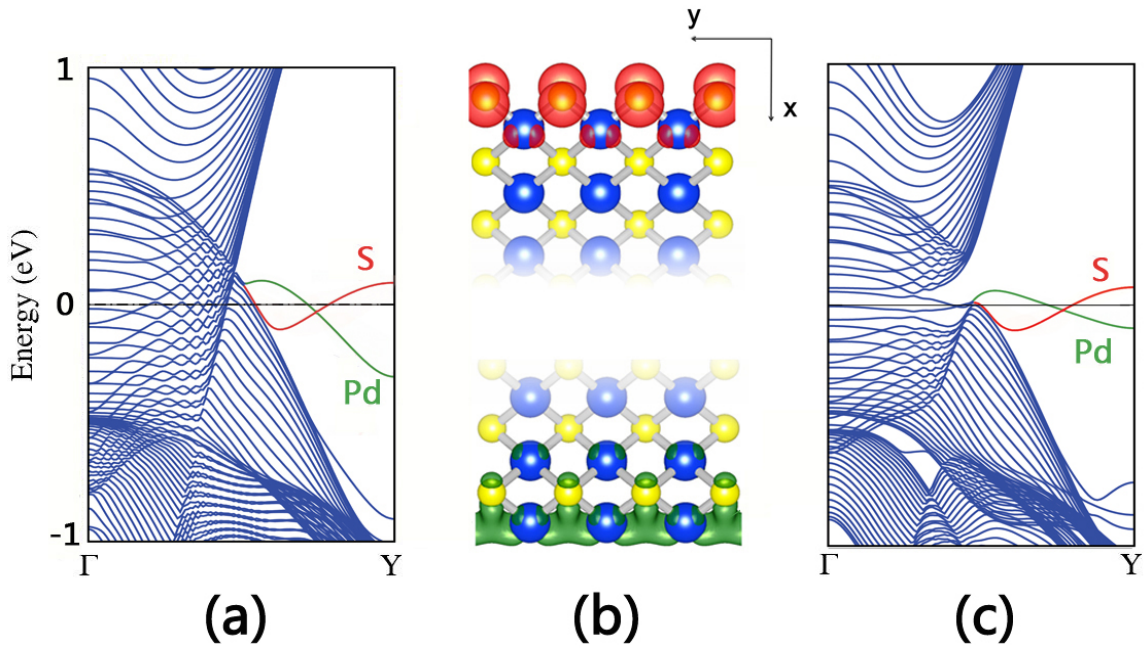

**Figure 4.** The electronic band structure of edge states of the ribbon (approximately 130 Å wide), measured in the *x* direction, which is calculated without SOC. Contributions from Pd and S edge atoms are indicated by red and green respectively. (a) Band structure together with edge states are shown. (b) Isosurface charge densities shown in real space that contribute to respective edge states are shown. (c) Band structure of edge states where the unit cell structure is modified by shifting one Pd atom by a displacement with *x*, *y* and z components of 0.1 Å

**Corresponding Author**

*E-mail: xuh@sustc.edu.cn (H.X.).

*E-mail: tong.sy@sustc.edu.cn (S.Y.T.)

(Y.J., R.W. and J.Z.) These authors contributed equally to this work.

■ ACKNOWLEDGMENTS

The authors thank Professor W. Q. Chen, J. S. Wu, F. Ye, H. Z. Lu, Dr. X. L. Yu and Mr. B. Wan for useful discussions. This work is supported by the National Natural Science Foundation of China (NSFC, Grant Nos.11204185, 11334003, 11404159 and 11304403).